\documentstyle[12pt,epsf,german]{article}

\advance\textheight by \topskip
\renewcommand{\baselinestretch}{1.2}
\renewcommand{\thefootnote}{\fnsymbol{footnote}}
\newcommand{\beq}{\begin{equation}}      
\newcommand{\eeq}{\end{equation}}
\newcommand{\bea}{\begin{eqnarray}}
\newcommand{\eea}{\end{eqnarray}}

\renewcommand{\bar}[1]{\overline{#1}}
\def\slash#1{#1\!\!\!/\!\,\,}
\newcounter{hilf}

\begin{document}

\begin{titlepage}
  \renewcommand{\baselinestretch}{1}
  \renewcommand{\thefootnote}{\alph{footnote}}

  \thispagestyle{empty}
   {\bf \hfill                                       UWThPh--1997--01}

\vspace*{-0.3cm}
   {\bf \hfill                                       TUM--HEP--266/97/re}

\vspace*{-0.3cm}
   {\bf \hfill\hfill                                       January 1997} 

\vspace*{0.5cm} {\Large\bf
  \begin{center} SUSY Violation in Effective Theories  \end{center}}

\vspace*{1cm}  {\large\sc 
\begin{center} 
                Richard Dawid\footnote{\makebox[1.cm]{Email:}
                                  Richard.Dawid@merlin.pap.univie.ac.at}
               and Serguei Reznov\footnote{\makebox[1.cm]{Email:}
                                  Serguei.Reznov@Physik.TU-Muenchen.DE}

\vspace*{0cm} {\it \begin{center}
    \footnotemark[1]Institut f\"ur Theoretische Physik, Universit\"at
    Wien, \\ Boltzmanng. 5 A--1090 Wien, Austria

                   \ 

    \footnotemark[2]Institut f\"ur Theoretische 
    Physik, Technische Universit\"at M\"unchen,          \\
    James--Franck--Stra\ss e, D--85747 Garching, Germany
    \end{center} } 
\vspace*{2.3cm}               
 \end{center}}

                     {\Large \bf \begin{center} Abstract \end{center} } 

We show that the effective theory of a supersymmetric model can
violate SUSY at the level of dimension six operators and higher. 
This phenomenon occurs in gauge theories which involve heavy 
vector--superfields and different mass scales. It appears in SUSY GUT theories
and is important in models of propagating Higgs boundstates.

\renewcommand{\baselinestretch}{1.2}
\end{titlepage}

\newpage
\renewcommand{\thefootnote}{\arabic{footnote}}
\setcounter{footnote}{0}



The calculation of effective operators in supersymmetric theories plays an
important role in the determination of the low energy implications of SUSY GUT
theories  and other supersymmetric high energy concepts \cite{eff}. 
In the following
we show in a toy model that the expansion in local effective operators 
extracted from a supersymmetric underlying theory can violate SUSY. 
This phenomenon will turn out to be relevant in SUSY GUTs and supersymmetric
models of Higgs boundstates. 


{\large \bf An example of effective SUSY violation:} 
Our toy model consists of the MSSM plus an extra heavy gauge vector field
$V_S$ which acquires its mass by spontaneous symmetry breaking in an
additional heavy Higgs sector. 
We calculate the effective operators at tree level the following way:
We write down the component Lagrangian
in the Wess--Zumino gauge.
We  extract the equations of motion from the full
Lagrangian to get ``constituent relations'' for the heavy fields. These
relations
also include suppressed derivative terms of heavy fields
coming from their kinetic terms. 
We insert these relations to eliminate 
the heavy fields in lowest order. Then we re--insert the same relations
again to eliminate the suppressed derivative  terms of heavy fields. 
This gives us the correct effective theory up to order $1/M_S^2$.
The part of the Lagrangian which involves heavy fields has the form

\beq
{\it L}_{heavy}=\int d^{2}\theta d^{2}\bar{\theta}
(\bar{Q}e^{(g_SV_{S}+g_2V_2+g_{Q}V_1)}Q+
Re^{(g_SV_{S}+g_{R}V_1)}\bar{R}+
H_Se^{(g_SV_{S}+g_{H_S}V_1)}\bar{H_S})+ Pot(H_S)
\label{fit}
\eeq

Throughout this paper superfields will always 
be denoted by capital letters  and 
component fields by small letters except for the vectorfield
which is identifiable by its Dirac index.
$Q$ and $R$ are the left-handed and right-handed quark--superfields
respectively, $H_S$
is the heavy Higgs, $V_S$ is the heavy vector--superfield, $V_2$ and $V_1$
are the SU(2) and U(1) MSSM gauge--superfields respectively. $Pot(H_S)$
denotes the operators which produce a scalar potential that breaks the
$V_S$ gauge symmetry in a supersymmetric way. One could  use a negative 
Fayet--Iliopoulos term to achieve this. However it is not necessary to
specify $Pot(H_S)$ because the heavy Higgses do not couple to light fields 
which has the consequence that $Pot(H_S)$ does not contribute to the effective
Lagrangian up to order $1/M^2$. 
Working in the Wess--Zumino(WZ) gauge the non-physical low-$\theta$ components
of the gauge superfield are gauged away and eq.~(\ref{fit}) 
produces, among others, the component terms

\beq
{\it L}_{heavy}=
\tilde{q}^{\dagger}g_SV_S^{\mu}(g_2{V_2}_{\mu}+g_{Q}{V_1}_{\mu})\tilde{q}+
\tilde{r}^{\dagger}g_SV_S^{\mu}(g_{R}{V_1}_{\mu})\tilde{r}+
\bar{q}\gamma^{\mu}g_S{V_S}_{\mu}q +
\bar{r}\gamma^{\mu}g_S{V_S}_{\mu}r+....,
\label{qqvv}
\eeq

where the the small letters always denote the components of the superfield
represented by the corresponding
capital letter.
They lead to the following effective terms involving one light vector field
and 2 left and right--handed (s)quarks:

\bea
{\it L}_{Veff}
&=&g_S^2\tilde{q}^{\dagger}(g_2{V_2}_{\mu}+g_{Q}{V_1}_{\mu})\tilde{q}
\bar{r}\gamma^{\mu}r +
g_S^2\tilde{r}^{\dagger}(g_{R}{V_1}_{\mu})\tilde{r}
\bar{q}\gamma^{\mu}q \nonumber \\  &+& 
g_S^2\tilde{q}^{\dagger}(g_2{V_2}_{\mu}+g_{Q}{V_1}_{\mu})\tilde{q}
\tilde{r}^{\dagger}\slash{\partial}\tilde{r} +
g_S^2\tilde{r}^{\dagger}(g_{R}{V_1}_{\mu})\tilde{r}
\tilde{q}^{\dagger}\slash{\partial}\tilde{q}
\label{qqqqg}
\eea

Now the requirement of a supersymmetric structure of the effective theory 
would imply that each contribution in (\ref{qqqqg}) is part of a 
superfield D--term. (F--terms
cannot include the gauge boson.)
The corresponding D--term is

\beq
\int d^{2}\theta d^{2}\bar{\theta}
g_S^2\bar{Q}(g_2V_2+(g_{Q}+g_{R})V_1)Q\bar{R}^{c}R^{c}
\label{qqqqvsu}
\eeq

which, besides (\ref{qqqqg}) and similar terms, includes the component terms 

\bea
& &g_S^2\bar{q}(g_2{V_2}+g_{Q}{V_1})\tilde{q}\tilde{r}^{\dagger}r+
g_S^2\tilde{r}^{\dagger}(g_2{V_2}_{\mu}+g_{Q}{V_1}_{\mu})\tilde{r}
\bar{q}\gamma^{\mu}q +
g_S^2\tilde{q}^{\dagger}(g_{R}{V_1}_{\mu})\tilde{q}
\bar{r}\gamma^{\mu}r 
\label{qqqqgi3}\\
&+&(g_S^2\bar{q}(g_2{\lambda_2}+g_{Q}{\lambda_1})\tilde{q}\tilde{r}^{\dagger}
\tilde{r}+(q\leftrightarrow r)  \label{qqqqgi}\\
&+&g_S^2\tilde{q}^{\dagger}
(g_2D_{V_2}+g_{Q}D_{V_1})\tilde{q}\tilde{r}^{\dagger}
\tilde{r}  \label{qqqqgi2}~.
\eea

The terms of (\ref{qqqqgi3}), (\ref{qqqqgi})
and (\ref{qqqqgi2}) would be necessary to achieve a supersymmetric structure 
of our effective theory.
But, because of the
missing low-$\theta$ components of the vector superfield in the
WZ--gauge, they  are not produced by integrating out the heavy sector. 
This can be most obviously seen in the case of (\ref{qqqqgi})[see Fig. 1]. 
The light gauginos cannot couple to the heavy
sector: Couplings to heavy gauge bosons are excluded because they
belong to a different simple gauge group and couplings to heavy
scalars would produce large gaugino masses. Thus a local effective
coupling to the light gauginos could only come from covariant
kinetic terms of squarks. But in the WZ--gauge the contributions of covariant
kinetic terms that produce effective four--coupling terms have to include
heavy gauge bosons, heavy gauginos or D--fields. 
As the $\theta$--structure forbids 
to couple a  gaugino plus a second component of a vector superfield to the
quark sector in one gauge coupling term, the couplings of type 
(\ref{qqqqgi}) cannot exist in the effective theory.

\newpage 

\vskip -16cm
\begin{figure}[ht]
\epsfysize=18cm
\epsffile{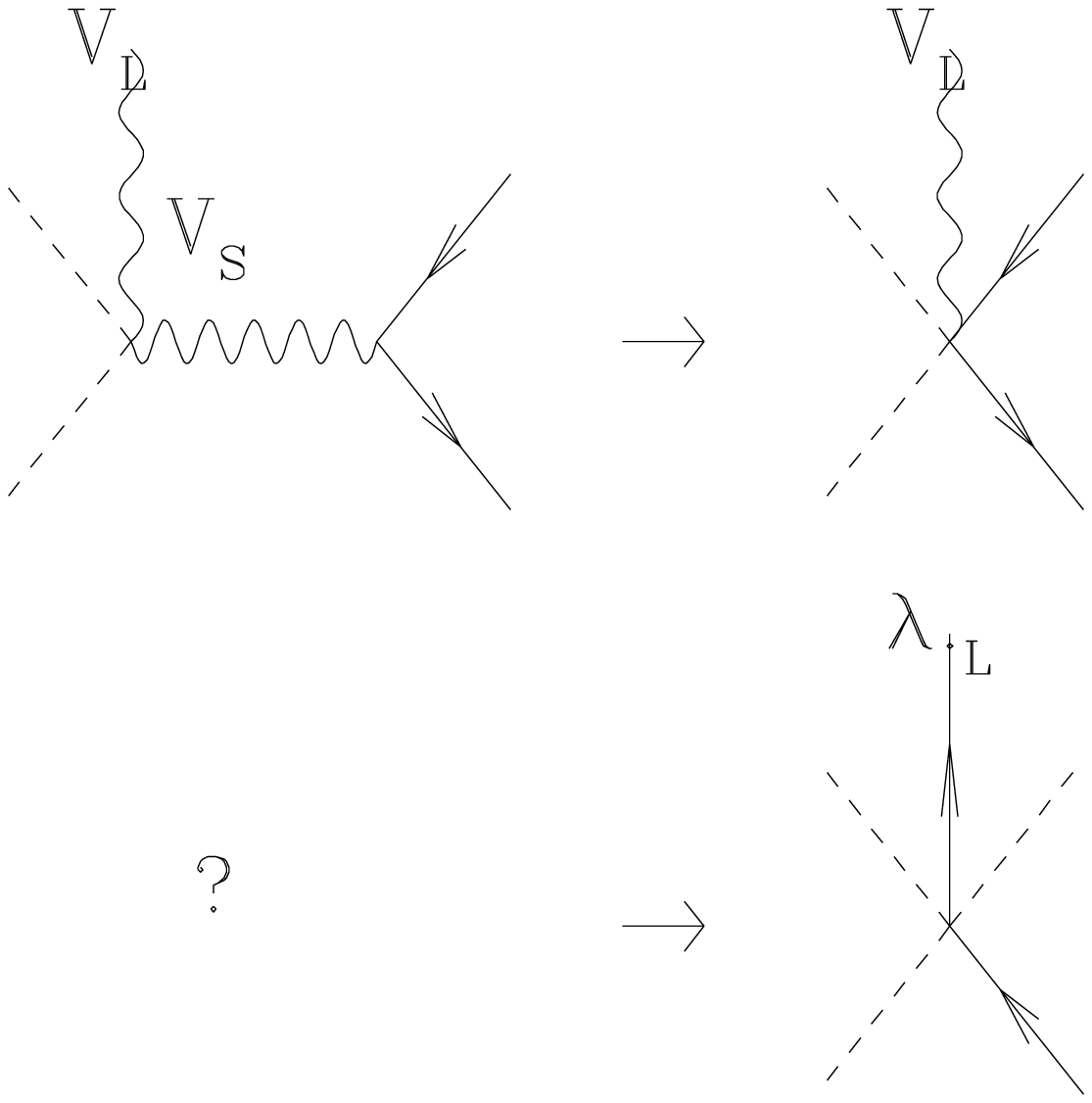}
\end{figure}
\vskip -6cm

{\it Fig.~1: SUSY violation due to missing light gaugino couplings. 
The effective diagrams on the right 
side belong to the same superfield coupling term. While the first one 
can be derived from a heavy gauge field exchange, the second one has
no correspondence in the full theory. The outer lines denote (s)quark fields.
L=1,2.}

\vspace{0.5cm}

A similar phenomenon already occurs if one neglects the light gauge fields
but takes into consideration the light Yukawa couplings.
In this case the heavy Lagrangian has the form

\beq
{\it L}^{\prime}_{heavy}=\int d^{2}\theta d^{2}\bar{\theta}
(\bar{Q}e^{g_SV_{S}}Q+
R^{c}e^{g_SV_{S}}\bar{R}^{c}+
H_S^{c}e^{g_SV_{S}}\bar{H_S}^{c})+ Pot(H_S) +\int d^{2}\theta g_yHQR~.
\label{fit2}
\eeq

The effective theory now includes most but not all component contributions
of the four--superfield operator

\beq
\frac{g_S^2}{M_{S}^2}
\int d^{2}\theta d^{2}\bar{\theta}[(\bar{Q}~\bar{R}^{c})
(QR^{c})]~. \nonumber
\label{qqqqsup}
\eeq

The couplings to quark auxiliary fields 
$F_Q, F_R$ are missing in the  
effective theory. The missing terms would for example correspond to a term
of the type
$1/M_S^2(squark)^2(quark)^2(light Higgs)$ after integrating out 
$D$-- and $F$--fields. This term once again cannot be produced by 
integrating out heavy gauge bosons. 

Only in a model which lacks both, light Yukawa terms and light gauge 
fields, SUSY is not violated in the effective theory. 
The lowest auxiliary field
contributions now are of order $M^{-4}$ and correspond to two--$V_S$
exchange in the full theory. Thus the term (\ref{qqqqsup}) seems to be the
correct effective operator in this case. The supersymmetric structure of
the effective operators in this special case seems to be somewhat connected
to the infinite increase of the number of fields which locally coupled
in increasing order $1/M_S$ in the effective theory.
However we did not find a satisfying structural 
explanation for the surviving of SUSY here.

The discussion above leads to the rather surprising 
conclusion that the low energy effective approximation of a supersymmetric 
theory can violate SUSY.
In the described cases the supersymmetric structure of the 
dimension six four--superfield coupling
cannot be achieved by integrating out heavy gauge bosons.


{\large \bf A characterization of the phenomenon:}
The phenomenon is connected to the special nature of the WZ--gauge. The 
WZ--gauge is not manifestly supersymmetric, it is not preserved by
SUSY transformations.
In a full gauge theory this is merely a formal problem.
One can modify the SUSY transformation by adding a
supersymmetric gauge transformation so that this modified SUSY transformation
preserves the WZ--gauge.
However the situation is different in an effective low energy approximation
of this gauge theory: After the heavy vector superfield
has been integrated out
the effective theory is not gauge invariant anymore and 
the modified SUSY transformation cannot be applied.
Therefore an effective theory gained from integrating out vector
superfields in the WZ--gauge is not a supersymmetric theory.
Now at first sight the conclusion could be tempting that it is illegal
to integrate out in the WZ--gauge. Integrating out in a general gauge
would give a supersymmetric result and the problem seems to be solved.
We argue however that the opposite is true and the
 mistake would be made by {\em not} using the
WZ--gauge: As a matter of
fact gauge invariance of the full theory makes some of the vector
superfield components non--physical. The non--physical character
of these fields cannot be seen anymore in the non--gauge--invariant
effective theory because the symmetry transformations to rotate them away
are no more defined. Therefore, if one would integrate out in a general
gauge, one would incorrectly believe in degrees of freedom which are
not there. The only way to avoid this is to use up the gauge 
freedom already in the full theory to make all non--physical fields
vanish. This means that it is necessary to use the WZ--gauge for
integrating out heavy degrees of freedom. But that can imply SUSY violation
in the resulting effective theory. 

It is instructive to have a short look at a situation where the low--$\theta$
components of the vector supermultiplet are physical. This is the case
if we replace the heavy Higgs sector by an explicit gauge symmetry breaking 
mass term $\int d^{2}\theta d^{2}\bar{\theta} m^2V_S^2$. 
This term corresponds to the component operators

\beq
\int d^{2}\theta d^{2}\bar{\theta} m^2V_S^2=
mCD + m(\bar{\chi}\lambda+\bar{\lambda}\chi) +M^{\dagger}M +N^{\dagger}N
+m^2V^2 +\bar{\chi}\slash{\partial}\chi +(\partial C)^2~,
\nonumber
\label{Vmass}
\eeq

with a scalar $C$, a fermionic $\chi$ and the auxiliary fields $M$ and $N$.
In this case there is no gauge freedom, the low--$\theta$ components
$\chi$ and $C$ get a kinetic term from the superfield mass term defined in
eq.~(\ref{Vmass}).
Integrating out $V_S$ in this model leads to a 
supersymmetric effective coupling. Argued in the framework of the discussion
above the effective operators must be supersymmetric because the `trick'
with the WZ--gauge cannot be applied. From a different, maybe slightly more 
physical point of view one can argue the following way:
If we have an explicit mass term, the
propagation of the heavy fields could be set equal to zero except for the
low--$\theta$ components which in this case come up to the derivative
couplings of the effective coupling term. $V_S$ can then be seen as something
like a vector--like auxiliary field and the $V_S$ exchange becomes a 
supersymmetric local operator. Now if we turn on again the kinetic term,
this gives just contributions of higher order and  cannot harm the 
supersymmetric effective dimension 6 operator any more.
SUSY is protected because the propagation of the heavy fields is no
essential feature of the model.

However a full gauge theory is a very different theory 
and it is exactly this difference that is overlooked by integrating
out in a general gauge.
In a gauge theory where the mass is provided by
spontaneous symmetry breaking via the vev of a Higgs field, 
the propagation of this Higgs is necessary for the mass term which 
is of course necessary itself for the procedure of integrating out. 
Therefore the nonlocal character is essential in this case and
SUSY which is no inner symmetry and therefore not blind against
propagation is not protected in an effective theory.

It is important to notice the following subtlety in the nature of effective
SUSY violation:
The full $1/M_S$--expansion of the fundamental theory
gives an infinite sum of non--supersymmetric local operators that
produce a supersymmetric theory (= the fundamental theory). 
By neglecting operators of orders higher 
than e. g.  $1/M_S^2$ one violates SUSY. Therefore the actual SUSY violation
is of order $1/M_S^3$.
Nevertheless the mistake one would make by misjudging the dimension
six operators as supersymmetric is of the order $1/M_S^2$ according
to the discussion of the previous section. A supersymmetric structure of
the dimension six operators would require additional operators of that
dimension which are simply missing. This means that
there exist two similar but different supersymmetric theories, one being
the theory of supersymmetric dimension six operators and the other being
the full theory of heavy gauge superfield exchange. A low energy effective
theory that wants to be correct up to order $1/M_S$ can differ from the second
only at order $1/M_S^3$ but differs from the first at order $1/M_S^2$. 


{\large \bf Relevance of effective SUSY violation:}
In our toy model we observed SUSY violating operators of dimension six.
One can easily understand that dimension four and five operators cannot
show the same phenomenon. The SUSY violating effect is intimately linked
to supersymmetric gauge invariance and therefore only occurs in the
supersymmetric sector. But in a scenario of vector superfield exchange
effective operators of dimension four and 
five only stem from the soft breaking sector and thus are not endangered
by the described SUSY violating mechanism.

On the level of dimension six effective operators however SUSY violation
is a general feature of every model with heavy gauge vector superfield
exchange plus couplings that involve just light fields. (The second condition 
is necessary to exclude the case of a pure heavy gauge field exchange which,
as we argued above, is supersymmetric.) 

One model where this SUSY 
violating phenomenon is of special importance is the model of propagating
Higgs boundstates like it is realized in supersymmetric top condensation
\cite{SBHL}. In a recent paper \cite{DR1} we argued that 
the non--renormalizable dimension six four--superfield operator
which is introduced in this model to achieve dynamical electroweak 
symmetry breaking cannot be interpreted as an effective operator
for heavy gauge superfield exchange. Now we find one more argument against
this interpretation: By neglecting the light Higgs sector 
of our toy model and adding some soft breaking terms 
we get exactly the type of heavy gauge theory which had to be
considered as the underlying gauge theory of SUSY top condensation. However we 
saw that this theory does not produce supersymmetric dimension six operators
and therefore is fundamentally incapable of leading to SUSY top condensation.

An important field where effective operators of a high scale SUSY theory
play a role are SUSY GUT theories. Here the SUSY violating effects also
occur albeit not on a level relevant for the low energy effects 
currently discussed.
The only difference between a SUSY GUT and our toy model concerns
the gauge structure: While in the toy model we had separate gauge groups
for light and heavy vector bosons now both sit in the adjoint representation
of one simple group. This different gauge structure implies some 
differences, if we 
reconsider our procedure of integrating out heavy component fields in the 
WZ--gauge.  Concerning for example the light gaugino case now there 
exist effective operators involving the
light gaugino $\lambda_L$. These stem from coupling terms of the form

\beq 
g_{gauge}V_S\bar{\lambda}_L\lambda_S
\label{lgi}
\eeq

in the full theory. However these effective operators are of dimension 7
and fail to produce a supersymmetric structure of dimension six operators.
The different gauge structure does not change the principle of the
discussed SUSY violating effect. 

\vskip -4cm
\begin{figure}[ht]
\epsfysize=18cm
\epsffile{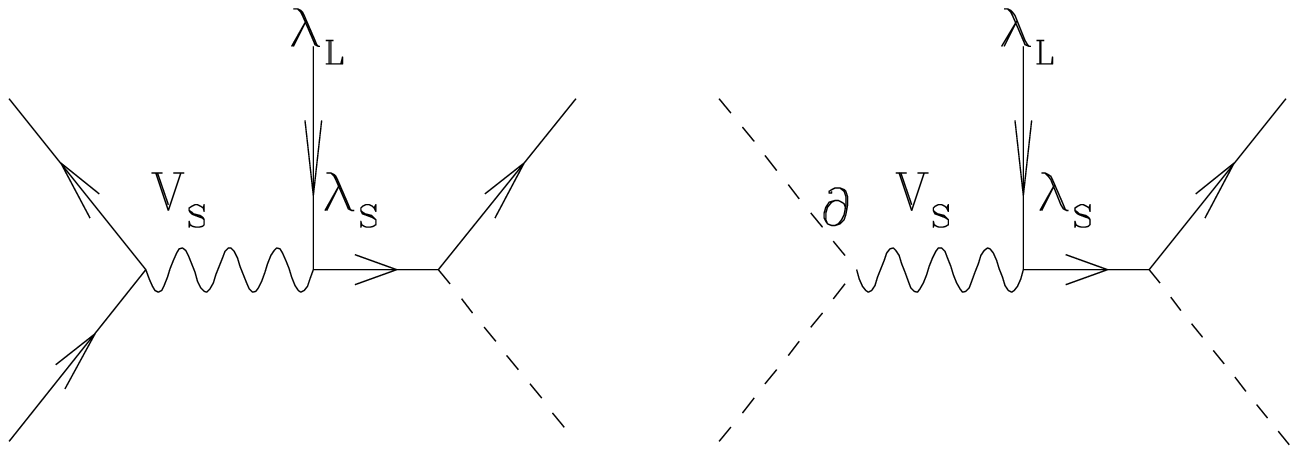}
\end{figure}
\vskip -10cm

{\it Fig.~2: Diagrams leading to effective dimension 7 operators with a 
light gaugino in SUSY GUTs. The outer lines again denote (s)quark fields.}

\vspace{0.5cm}

In a recent work \cite{dimpo} soft breaking GUT
contributions have been calculated by integrating out
the superfields in superspace which per definition must always lead to a
supersymmetric result. This comes up to integrating out in components
in a general gauge, as there is no gauge defined before expanding in
components. The method is perfectly valid if one knows in advance that
no SUSY violation can occur in the effective terms one wants to calculate.
This is the case for dimension 4 and 
dimension 5 operators like those calculated in \cite{dimpo}: 
In general one would have to be careful. 

A generally correct method would be the following: \\
- Calculate the equations of motion in superfields. \\
- If just chiral fields are integrated out, SUSY is safe and the whole
discussion can be done in superfields. \\
- If vector superfields are to be integrated out,
expand the equations of motion into component fields, write the
vector superfields in the WZ--gauge 
and carry out a comparison of coefficients. \\
- The result are constituent relations for the high $\theta$--valued
components of the vector superfield plus conditions which force
the low $\theta$--valued components of the supercurrent that couples to
the vector superfield to be zero. \\
-  Inserting the constituent conditions and applying the zero conditions
one gets the correct low energy effective theory.


{\large \bf Conclusions:} 
We described the possibility of SUSY violating effects in non--renormalizable
low energy effective theories of supersymmetric models. These effects
occur in any fundamental theory with heavy gauge superfields plus
any type of Yukawa or gauge coupling between light fields. 
Examples for this scenario are SUSY GUT theories or attempts to produce
SUSY top condensation by heavy vector exchange. SUSY violation happens
first at the level of dimension six operators where some of the 
component operators which would be necessary
to provide a supersymmetric structure at this level are not produced
in the effective theory. The foundation of the
phenomenon lies in the specific character of supersymmetric gauge
invariance which is lost in the effective theory. While the missing operators
are of order $1/M_S^2$ the SUSY
violating effect is never\-the\-less of higher order because the inclusion
of the  neglected higher
order operators would completely define the full theory and therefore
also reinstall SUSY. Still the mistake one would make
by constructing the effective theory supersymmetrically would be of
order $1/M_S^2$. 
By integrating out either in components
in a general gauge or in superfields one is
insensitive to effective SUSY violation and gets
wrong results if this phenomenon occurs.  
It is therefore necessary to integrate out in the WZ--gauge. 
The most convenient method is to extract the equations of motion in 
superfield formalism and fix the WZ--gauge by doing a 
comparison of coefficients in the
equations of motion before reinserting them into the Lagrangian.

\vspace{.5cm}
{\bf Acknowledgments:} We would like to thank
A. Blumhofer, D. Kominis, M. Lindner, M. Moser, H. P. Nilles, and  M. Yamaguchi
for useful discussions and helpful comments
on the draft version of this paper.

This work is in part supported by DFG.



\newpage
\vspace{1.cm}

\begin{thebibliography}{50}
\bibitem{eff}    S.~Weinberg, Phys.~Rev.~D26 (1982) 287;\\
                 N.~Sakai and T.~Yanagida, Nucl.~Phys.~B197 (1982) 533;\\
                 S.~Dimopoulos, S.~Raby and F.~Wilczek, Phys.~Lett.~112B
                 (1982) 133;\\
                 R.~Barbieri and R.~Gatto, Phys.~Lett. 110B (1982) 211;\\
                 J.~Ellis and D.~V.~Nanopoulos, Phys.~Lett.~110B (1982) 44;\\
                 T.~Inami and C.~S.~Lim, Nucl.~Phys.~B207 (1982) 533;\\
                 W.~Lang, Nucl.~Phys.~B203 (1982) 277;\\
                 L.~Hall, J.~Lykken and S.~Weinberg, Phys.~Rev.~D27 
                 (1983) 2359.
\bibitem{SBHL}   W.A.~Bardeen, T.E.~Clark and S.T.~Love, Phys. Lett. B237 
                 (1990) 235;\\
                 W.A.~Bardeen, M.~Carena, T.E.~Clark, K.~Sasaki and 
                 C.E.M.~Wagner, Phys. Lett. B369 (1992) 33. 
\bibitem{dimpo}  S.~Dimopoulos and A.~Pomarol, Nucl.~Phys.~B453 (1995) 83.
\bibitem{DR1}    R.~Dawid and S.~Reznov, Phys.~Lett.~B388 (1996) 315.

\end{thebibliography}
\end{document}